\documentclass[traditabstract,times]{aa} 

\newcommand{\ergps}{erg\thinspace s$^{-1}$}
\newcommand{\phpspsqcm}{ph\thinspace cm$^{-2}$\thinspace s$^{-1}$}
\newcommand{\ergpspsqcm}{erg\thinspace s$^{-1}$\thinspace cm$^{-2}$}
\newcommand{\psqcm}{cm$^{-2}$}
\newcommand{\nH}{$N_{\rm H}$}

\usepackage{graphicx}
\begin{document}
\title{The active nucleus of the ULIRG IRAS F00183--7111 viewed by NuSTAR\thanks{This research has also made use of data obtained from ESO telescopes at the La Silla Paranal Observatory. The ESO VLT data are under programme IDs 386.B-0346, 088.B-0405, and 090.B-0098.}}


   \author{K. Iwasawa\inst{1,2}
          \and
H.W.W. Spoon\inst{3}
\and
A. Comastri\inst{4}
\and
R. Gilli\inst{4}
\and
G. Lanzuisi\inst{4,5}
\and
E. Piconcelli\inst{6}
\and
C. Vignali\inst{4,5}
\and
M.~Brusa\inst{4,5}
\and
S. Puccetti\inst{7}
}

\institute{Institut de Ci\`encies del Cosmos (ICCUB), Universitat de Barcelona (IEEC-UB), Mart\'i i Franqu\`es, 1, 08028 Barcelona, Spain
         \and
ICREA, Pg. Llu\'is Companys 23, 08010 Barcelona, Spain
\and
Cornel University, Astronomy Department, Ithaca, NY 14853, USA
\and
INAF-Osservatorio Astronomico di Bologna, via Gobetti, 93/3, 40129 Bologna, Italy
\and
Dipartimento di Fisica e Astronomia, Universit\`a di Bologna, via Gobetti, 93/2, 40129 Bologna, Italy
\and
INAF-Osservatorio Astronomico di Roma, via Frascati 33, 00040, Monteporzio Catone, Italy
\and
Agenzia Spaziale Italiana-Unit\`a di Ricerca Scientifica, Via del Politecnico, 00133 Roma, Italy
          }


 
          \abstract{We present an X-ray study of the ultra-luminous
            infrared galaxy IRAS F00183--7111 ($z=0.327$), using data
            obtained from NuSTAR, Chandra X-ray Observatory, Suzaku
            and XMM-Newton. The Chandra imaging shows that a
            point-like X-ray source is located at the nucleus of the
            galaxy at energies above 2 keV. However, the point source
            resolves into diffuse emission at lower energies,
            extending to the east, where the extranuclear [O{\sc
                iii}]$\lambda 5007$ emission, presumably induced by a
            galactic-scale outflow, is present. The nuclear source is
            detected by NuSTAR up to the rest-frame 30 keV. The
            strong, high-ionization Fe K line, first seen by
            XMM-Newton, and subsequently by Suzaku and Chandra, is not
            detected in the NuSTAR data. The line flux appears to have
            been declining continuously between 2003 and 2016, while
            the continuum emission remained stable to within
            30\%. Further observations are needed to confirm this. The
            X-ray continuum below 10 keV is characterised by a hard
            spectrum caused by cold absorption of \nH $\sim 1\times
            10^{23}$ \psqcm, compatible to that of the silicate
            absorption at $9.7\mu $m, and a broad absorption feature
            around 8 keV which we attribute to a high-ionization Fe K
            absorption edge. The latter is best described by a
            blueshifted, high-ionization (log $\xi\sim 3$) absorber
            with a column density of \nH $\sim 1\times 10^{24}$
            \psqcm, similar to the X-ray high-velocity outflows
            observed in a number of active nuclei. No extra hard
            component, which would arise from a strongly absorbed
            (i.e. Compton-thick) source, is seen in the NuSTAR
            data. While a pure reflection scenario (with a totally
            hidden central source) is viable, direct emission from the
            central source of $L_{\rm 2-10 keV}\simeq 2\times 10^{44}$
            \ergps, behind layers of cold and hot absorbing gas may be
            an alternative explanation. In this case, the relative
            X-ray quietness ($L_{\rm x}/L_{\rm bol,AGN}\leq 6\times
            10^{-3}$), the high-ionization Fe line, strong outflows
            inferred from various observations, and other similarities
            to the well-studied ULIRG/QSO Mrk~231 point that the
            central source in this ULIRG might be accreting close to
            the Eddington limit.}

\titlerunning{NuSTAR and Chandra observations of IRAS F00183--7111}
\authorrunning{K. Iwasawa et al.}
   \maketitle
%

\section{Introduction}

IRAS F00183--7111 is an ultra-luminous infrared galaxy (ULIRG) with
log $L$(1-1000 $\mu $m) $= 9.9\times 10^{12}L_{\odot}$ (Spoon et al 2009) at $z
= 0.327$ and shows an archetypal absorption-dominated mid-infrared
spectrum with no polycyclic aromatic hydrocarbon (PAH) feature
detected at 6.2 $\mu$m (Spoon et al 2004). The silicate absorption
depth at 9.7 $\mu$m is however shallower than those in the branch of
deeply obscured sources (e.g. NGC 4418, IRAS F08572+3915) in the
``fork diagram'' of Spoon et al (2007). This apparent shallowness of
the silicate depth could be a result of dilution by the continuum which
leaks through ruptured absorbing screen, and it led Spoon et al
(2007, 2009) to suggest that IRAS F00183--7111 may be in the early
phase of disrupting the nuclear obscuration to evolve into the quasar
regime. The fast outflow signatures found in the mid-infrared lines
(Spoon et al 2009) and the VLBI-scale radio jets (Norris et al 2012)
appear to support this picture. The radio luminosity is in the range
of powerful radio galaxies and the radio excess with respect to the
infrared emission (Roy \& Norris 1997; Drake et al 2004) indicates
presence of a powerful active galactic nucleus (AGN). A XMM-Newton
observation detected a hard-spectrum X-ray source with a 2-10 keV
luminosity of $\sim 10^{44}$ \ergps. The X-ray spectrum shows a strong
Fe K feature, suggesting that a Compton thick AGN with a much larger
luminosity is hidden in this ULIRG (Nandra \& Iwasawa 2007; Ruiz,
Carrera \& Panessa 2007). The Fe K line is, however, found at the
rest-frame 6.7 keV, indicating Fe {\sc xxv}, i.e. highly ionized
line-emitting medium, which is unusual for an obscured AGN as it
normally shows a Fe K feature dominated by a 6.4 keV line from cold Fe
(less ionized than Fe {\sc xvii}).

The fast outflow signature of IRAS F00183--7111 was first identified
by the optical [O{\sc iii}]$\lambda 5007$ kinematics 
and the ionized gas extends by $\sim 10$ arcsec to the
east of the nucleus (Heckman, Armus \& Miley 1990). Much more enhanced
outflow signatures were found in the blueshifted mid-IR lines of
[Ne{\sc ii}]$\lambda 12.81 \mu$m and [Ne{\sc iii}]$\lambda 15.56 \mu$m
with velocity widths of FWZI$\sim 3000$ km s$^{-1}$ (Spoon et al 2009), which
presumably occur in the region close to the nucleus where dust
obscuration hides the optical signatures from our view. Weak soft
X-ray emission appears to be displaced from the hard X-ray position by
$\sim 5$ arcsec towards the east and is possibly associated with the
extended outflow structure.

We newly acquired X-ray data from Chandra X-ray Observatory (Chandra)
and NuSTAR to study further the X-ray properties of IRAS
F00183--7111. The arcsec resolution of Chandra imaging was used to
investigate the extended soft X-ray emission hinted by the XMM-Newton
observation. The hard X-ray spectrum obtained from NuSTAR was examined
for constraining the properties of the obscured active
nucleus. The Fe K emission is a key diagnostic for the nuclear
obscuration and the physical condition of the nuclear medium, for
which data of better quality are desired. The existing XMM-Newton data
and the Suzaku data from the public archive are supplemented to study
the Fe K band spectrum.

The cosmology adopted here is $H_0=70$ km s$^{-1}$ Mpc$^{-1}$,
$\Omega_{\Lambda}=0.72$, $\Omega_{\rm M}=0.28$ (Bennett et al
2013). For the redshift of IRAS F00183--7111 ($z=0.327$), the
luminosity distance is $D_{\rm L}= 1723$ Mpc and the angular-scale is
4.7 kpc arcsec$^{-1}$.

\section{Observations}

\begin{table*}
\begin{center}
\caption{Observation log of IRAS F00183--7111}
\begin{tabular}{lcccccc}
Observatory & Camera & Date & ObsID & Exposure & Count rate & (Band) \\
&&&& ks & ct/s & keV \\  
NuSTAR & FPMA/FPMB & 2015-12-21 & 6010105002 & 52.1 & $(2.5\pm 0.2)\times 10^{-3}$ & (3-24) \\
NuSTAR & FPMA/FPMB & 2016-04-26 & 6010105004 & 52.6 & $(2.5\pm 0.2)\times 10^{-3}$ & (3-24) \\
Chandra & ACIS-S & 2013-02-13 & 13919 & 22.8 & $(5.9\pm 0.5)\times 10^{-3}$ & (0.5-8) \\
Suzaku & XIS(0,1,3) & 2012-05-03 & 7070036010 & 94.0 & $(6.2\pm 0.4)\times 10^{-3}$ & (0.5-8) \\
XMM-Newton & EPIC(pn,MOS1,MOS2) & 2003-04-16 & 0147570101 & 8.4,11.4,11.4 & $(2.4\pm 0.2)\times 10^{-2}$ & (0.5-8) \\ 
\end{tabular}
\begin{list}{}{}
\item[] Note --- The NuSTAR count rates are measured with a single detector
  in the associated band. The Suzaku count rate shows a sum of the
  three detectors. The XIS2 has already ceased its operation when the
  observation was carried out and no data are available. The
  XMM-Newton count rate is the sum of the three EPIC cameras.
\end{list}
\end{center}
\end{table*}


X-ray observations of IRAS F00183--7111 with four X-ray observatories,
XMM-Newton, Chandra, Suzaku and NuSTAR are listed in Table 1. Our
NuSTAR data of a total exposure time of 105 ks were taken in two
occasions separated by four months. The data were calibrated and
cleaned, using the NuSTARDAS included in HEASOFT (v6.19). In addition
to the default data cleaning, some time intervals near the south
atlantic anomaly (SAA) passage with elevated background were discarded
by applying the screening with {\tt SAAMODE=optimized} and {\tt
  TENTACLE=yes}, which removed further 4 ks for the first
observation. On inspecting the 3-20 keV light curve obtained from the
whole field of view, short intervals ($\leq 2$ ks in total) of solar
flares were noted during the first observation, which may affect the
background subtraction around 10 keV (Lanzuisi et al 2016). However,
as no notable effect was found in the spectrum of the first
observation, the intervals were not excluded. The NuSTAR spectra
obtained from the two detector modules FPMA and FPMB and the two
observations agree with each other within error, and the four datasets
are combined together for the analysis below. The Chandra and XMM-Newton
data were reduced using the standard software CIAO (v4.8) and XMMSAS
(v15.0), respectively. The Suzaku data (PI: E. Nardini) were taken from
the public archive and reduced using XSELECT of HEASoft. All the
spectral data were analysed using the spectral analysis package XSPEC
(v12.9).

\section{Results}

\subsection{Chandra imaging}


\begin{figure}
  \centerline{\includegraphics[width=0.5\textwidth,angle=0]{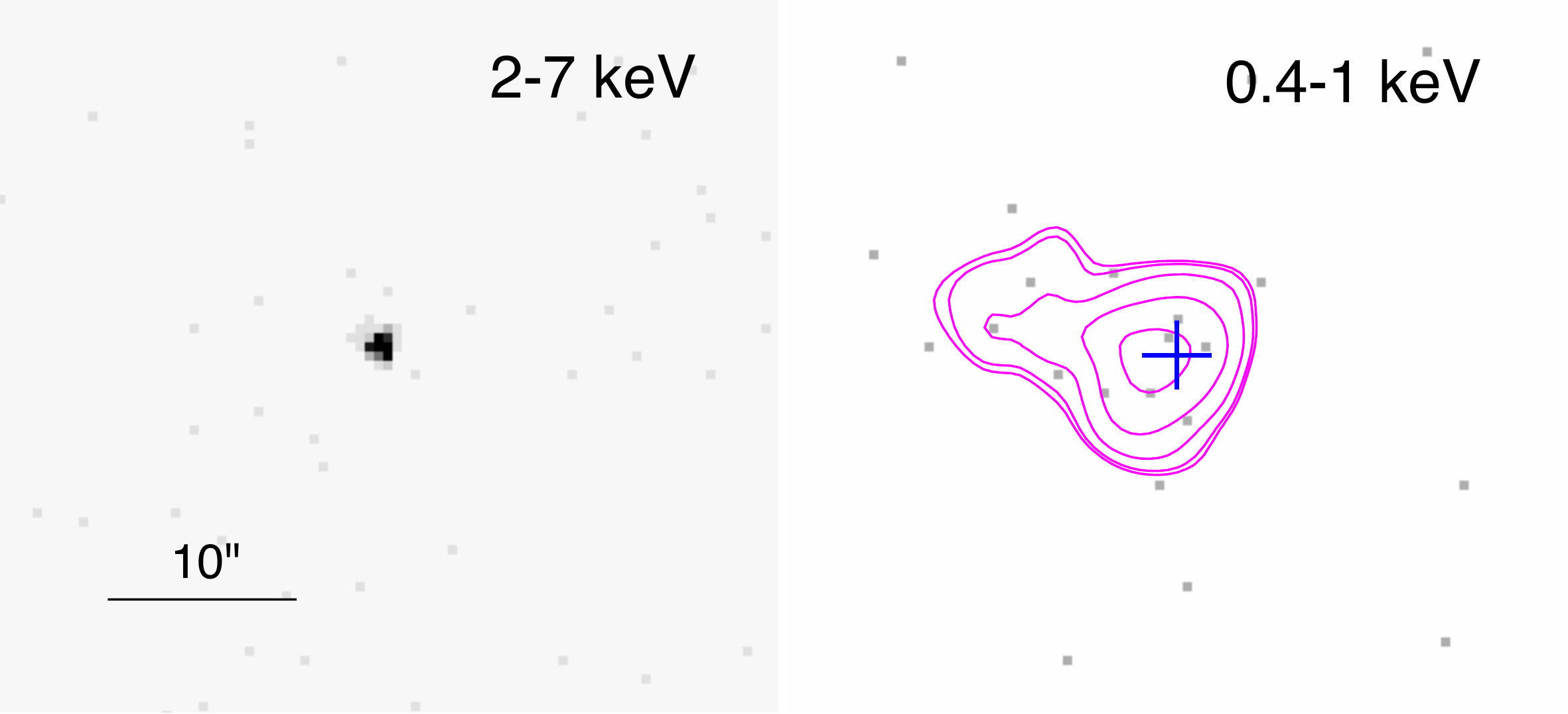}}
  \vspace{3mm}
  \centerline{\includegraphics[width=0.34\textwidth,angle=0]{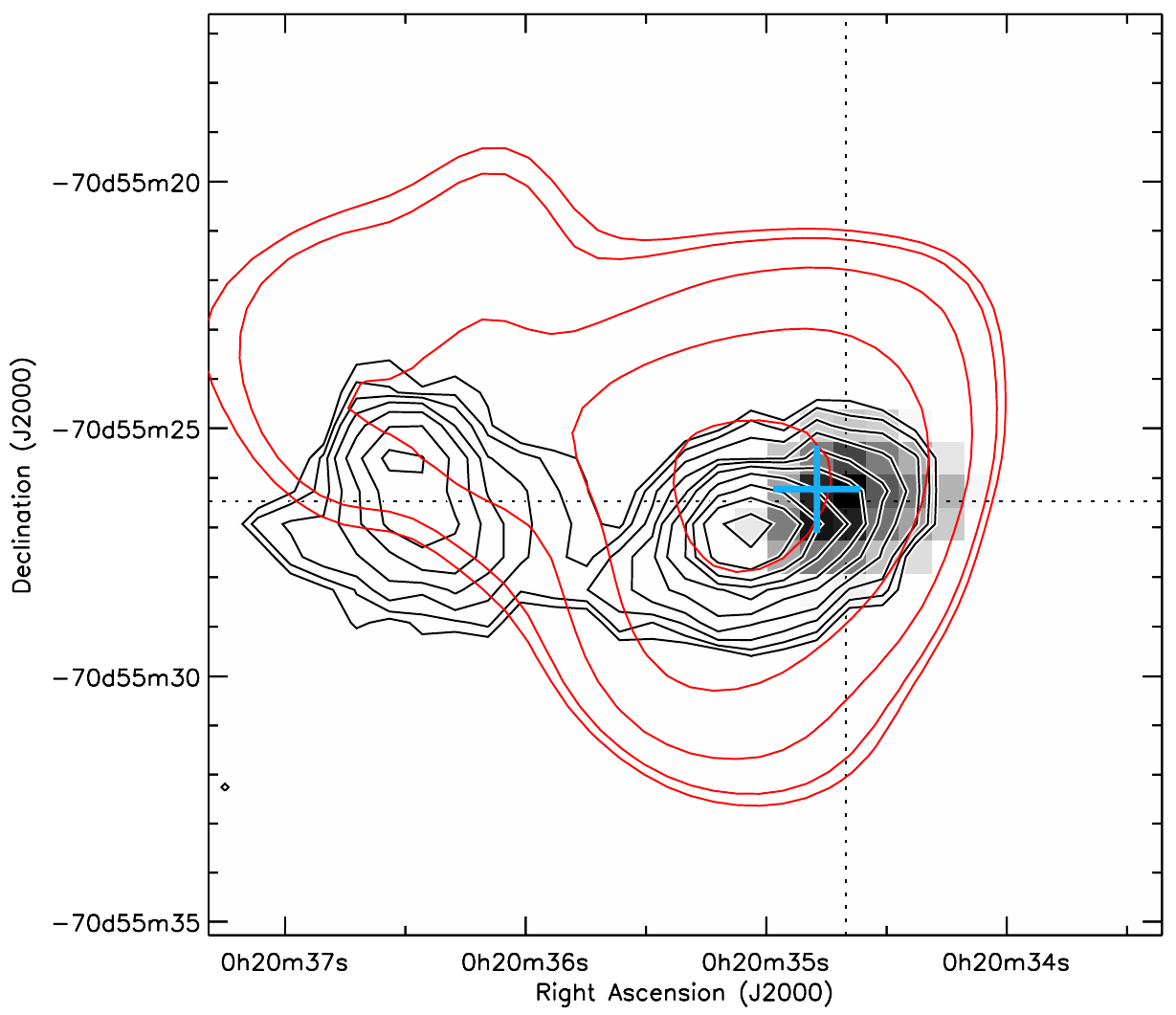}}
\caption{Upper-Left: a) The 2-7 keV Chandra image of IRAS
  F00183--7111. The X-ray source is point-like and coincides with the
  nuclear position of the ULIRG; Upper-right: b) The 0.4-1 keV Chandra
  image of the same area. The blue cross indicates the position of the
  2-7 keV source. The magenta contours for a smoothed version of the
  same image are overlaid. The image is elongated towards the eastern
  direction; Lower panel: c) The [O{\sc iii}]$\lambda 5007$ contours
  (black) are overlaid on the 5007\AA\ continuum in greyscale. The
  images were taken by VLT/VIMOS (Spoon et al, in prep.). The
  displacement between the [O{\sc iii}] and the continuum is
  considered to be due to obscuration in the nucleus, as discussed in
  Spoon et al (2009). The soft X-ray contours are also overlaid in
  red, and the 2-7 keV image centroid is marked in blue. An eastern
  [O{\sc iii}] knot is seen at 8 arsec from the nucleus.}
\end{figure}

The arcsecond resolution of the Chandra image localizes an X-ray
source at the position of the IRAS F00183--7111 nucleus (Fig 1a). The
source at energies above 2 keV is point-like. However, the point-like
source disappears below 1 keV, leaving only faint emission which
extends to the east up to $\sim 15$ arcsec, as suggested by the
overlaid contours (Fig. 1b). With the exposure time of 23 ks, 9 counts
are detected along this eastern extension in the 0.4-1 keV band
outside of the 2 arcsec from the nucleus position. When the annulus
of 2-18 arcsec from the nuclear position is divided azimuthally to
four quadrants, the eastern quadrant contains these 9 counts and the
other three contain 2 (northern), 3 (southern) and 1 (western). The
background counts are estimated to be $1.7\pm 0.1$ counts for each
quadrant. The eastern region has 5 times excess counts of the
background while the other regions have counts comparable to the
background.  While we can thus only remark that detected counts are
clustered on the eastern side of the nucleus, the small detected
counts do not warrant any further imaging analysis. The one-sided
extension is compatible with the eastern offset seen in the soft X-ray
image of XMM-Newton (Nandra \& Iwasawa 2007). The mean surface
brightness of this extended emission is $\sim 2\times 10^{-17}$
\ergpspsqcm\ arcsec$^{-2}$

It has been known that [O{\sc iii}]$\lambda 5007$ emission extends
towards the east in this ULIRG (Heckman, Armus \& Miley 1990). Our
recent VLT VIMOS observation reveals more details of the optical
extended nebula (Spoon et al. in prep.), and the [O{\sc iii}] contours
overlaid onto the continuum image, obtained from the VIMOS data, is
shown in Fig. 1c. Note that the scale of this image differs from the
X-ray images (a factor of $\sim 2$ smaller). It is unclear whether
this optical emission-line nebula is closely related to the extended
soft X-ray emission. They both extend towards the east and some soft
X-ray photons are detected along the [O{\sc iii}] extension. However,
the [O{\sc iii}] image shows a distinctive, bright knot at 8 arcsec
from the nucleus (with PA$\simeq 270^{\circ}$), which has no X-ray
counterpart. In contrast, the soft X-ray emission is diffuse and the
general direction of the extension may be slightly tilted towards
north (PA$\sim 250^{\circ}$) from that of [O{\sc iii}].

\subsection{X-ray spectrum}

We present results on the X-ray spectrum of IRAS F00183--7111 as
follows. The iron K emission line was investigated combining the data
from XMM-Newton, Chandra and Suzaku, as they have a comparable
spectral resolution (Sect. 3.2.1). Then the flux variability of the Fe
line (Sect. 3.2.1) and the underlying 2-8 keV continuum emission
(Sect. 3.2.2) between the observations was measured including the
NuSTAR data. The 2-20 keV continuum spectrum can be described by an
absorbed power-law and the cold absorbing column is constrained well,
combining the NuSTAR data with the low-energy data from the other
instruments. Then the broad-band NuSTAR spectrum was investigated for
the hard band spectral properties (Sect. 3.2.3).

\subsubsection{Fe K line}


\begin{figure}
\centerline{\includegraphics[width=0.4\textwidth,angle=0]{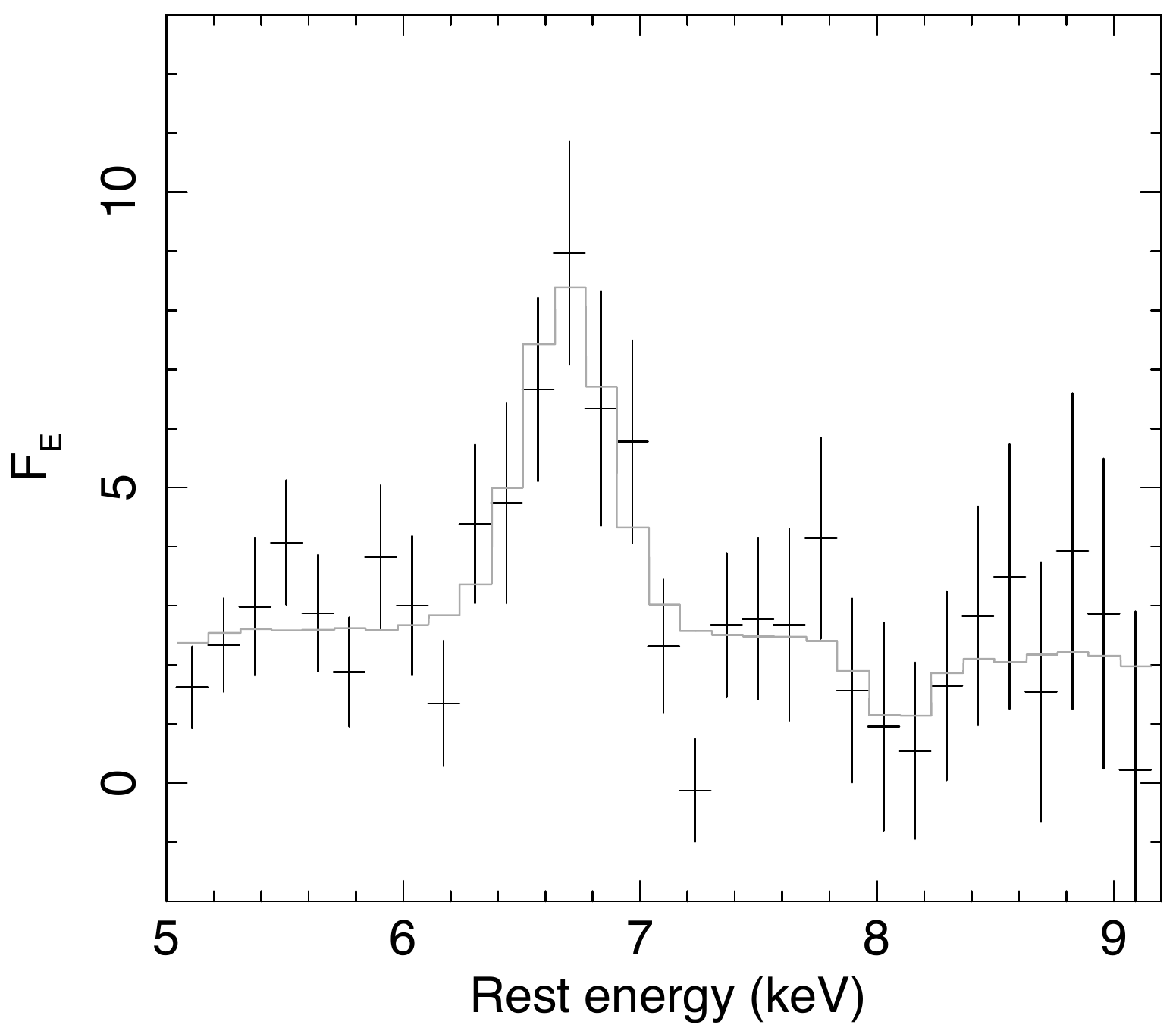}}
\caption{The rest-frame Fe K band spectrum of IRAS F00183--7111. Data
  obtained from XMM-Newton EPIC cameras, Chandra ACIS-S, Suzaku XIS,
  were combined. The data are in flux density in units of $10^{-14}$
  erg s$^{-1}$ cm$^{-2}$ keV$^{-1}$. The Fe line feature peaks at 6.7
  keV and has a moderate width of $\sigma\sim 0.15$ keV. For
  reference, the blueshifted absorption model for the NuSTAR spectrum
  discussed in Sect. 3.2.3 (IABS2 in Table 2), supplemented by a
  Gaussian line for the Fe K feature, is overplotted by the histogram
  in grey. The continuum level is corrected for the flux difference
  (see Sect. 3.2.2). The blueshifted Fe {\sc xxv} absorption line at 8
  keV predicted by the absorption model seems to be matched by the
  data.}
\end{figure}

A line feature at 5 keV was detected by the XMM-Newton observation and
identified with an iron K line of Fe {\sc xxv} at the rest-energy of
6.7 keV (Nandra \& Iwasawa 2007; Ruiz et al 2007). We inspected the
presence of the Fe K line by further adding the data from the Chandra
ACIS-S and the Suzaku XIS, which have similar spectral resolutions to
that of the XMM-Newton EPIC cameras. The 3.7-7 keV (5-9 keV in the
rest-frame) spectrum of a 100-eV resolution, obtained by combining
XMM-Newton, Chandra and Suzaku, after correcting for the respective
detector responses is shown in Fig. 2 (note that this plot is only for
displaying purposes). The plotted data were averaged with weighting by
the signal to noise ratio. The line is now detected at $4 \sigma $
significance above the neighbouring continuum in the 3.7-7 keV range
with average intensity of $(2.1\pm 0.5)\times 10^{-6}$
\phpspsqcm. Fitting a Gaussian to the line feature in all the spectral
data jointly gives the centroid energy of $6.69\pm 0.04$ keV in the
galaxy rest-frame. The line feature is resolved: $\sigma = 0.16\pm
0.03$ keV in Gaussian dispersion, which is significantly broader than
that expected from a Fe {\sc xxv} complex ($\sigma\sim 0.02$ keV). The
calibration errors in energy scale for respective instruments are at
the level of $<10$ eV (M. Guainazzi, priv. comm.), which cannot
account for the broadening. The line detection and the above results
on the line parameters are robust against continuum modelling (see
below).

The equivalent width of the line feature with respect to the local
continuum is $1.3\pm 0.3$ keV (the mean of the three observations,
corrected for the galaxy redshift). No clear 6.4 keV cold Fe K line is
seen. The 90\% upper limit on a narrow line at rest-frame 6.4 keV is
0.16 keV.

\begin{figure}
\centerline{\includegraphics[width=0.5\textwidth,angle=0]{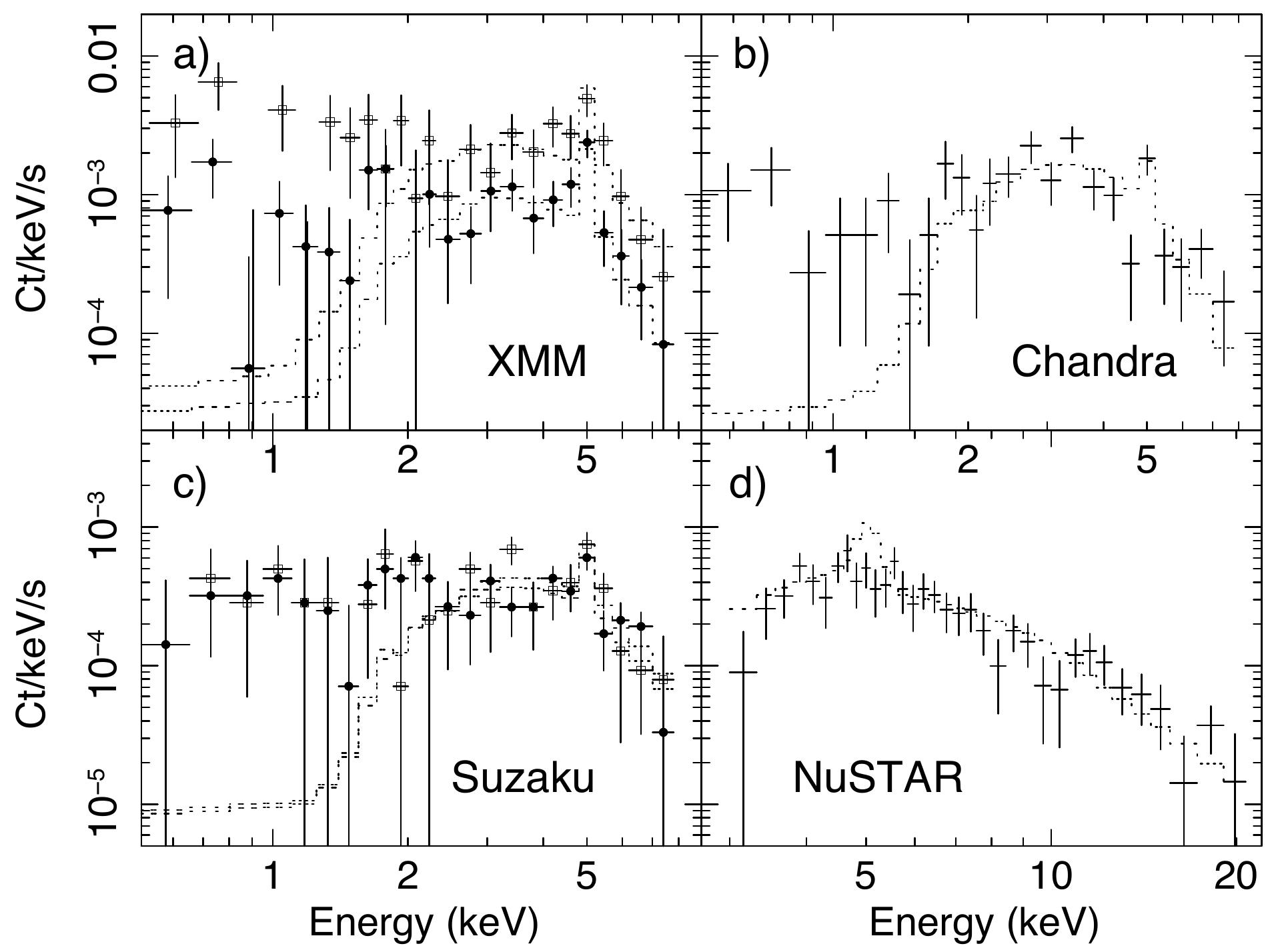}}
\caption{Energy spectra of IRAS F00183--7111, observed with four X-ray
  observatories: a) XMM-Newton: EPIC pn (open squares) and EPIC MOS1
  and MOS2 combined (solid circles); b) Chandra ACIS-S; c) Suzaku XIS0
  (solid circles); XIS1 and XIS3 combined (open squares); and d)
  NuSTAR: FPMA and FPMB combined. The dotted-line histogram in each
  panel indicates the best-fit model of an absorbed power-law plus a
  Gaussian for the 2-20 keV data obtained by fitting to all the
  datasets jointly (see text). Spectral parameters are identical
  between the observatories apart from the normalizations of the
  power-law and the Gaussian line. Emission below 2 keV originates
  from an extended extranuclear region (Fig. 1). Note that the model
  for Fe K in d) shows a line feature with its intensity observed with
  XMM-Newton for a comparison with the data. }

\end{figure}


\begin{figure}
\centerline{\includegraphics[width=0.4\textwidth,angle=0]{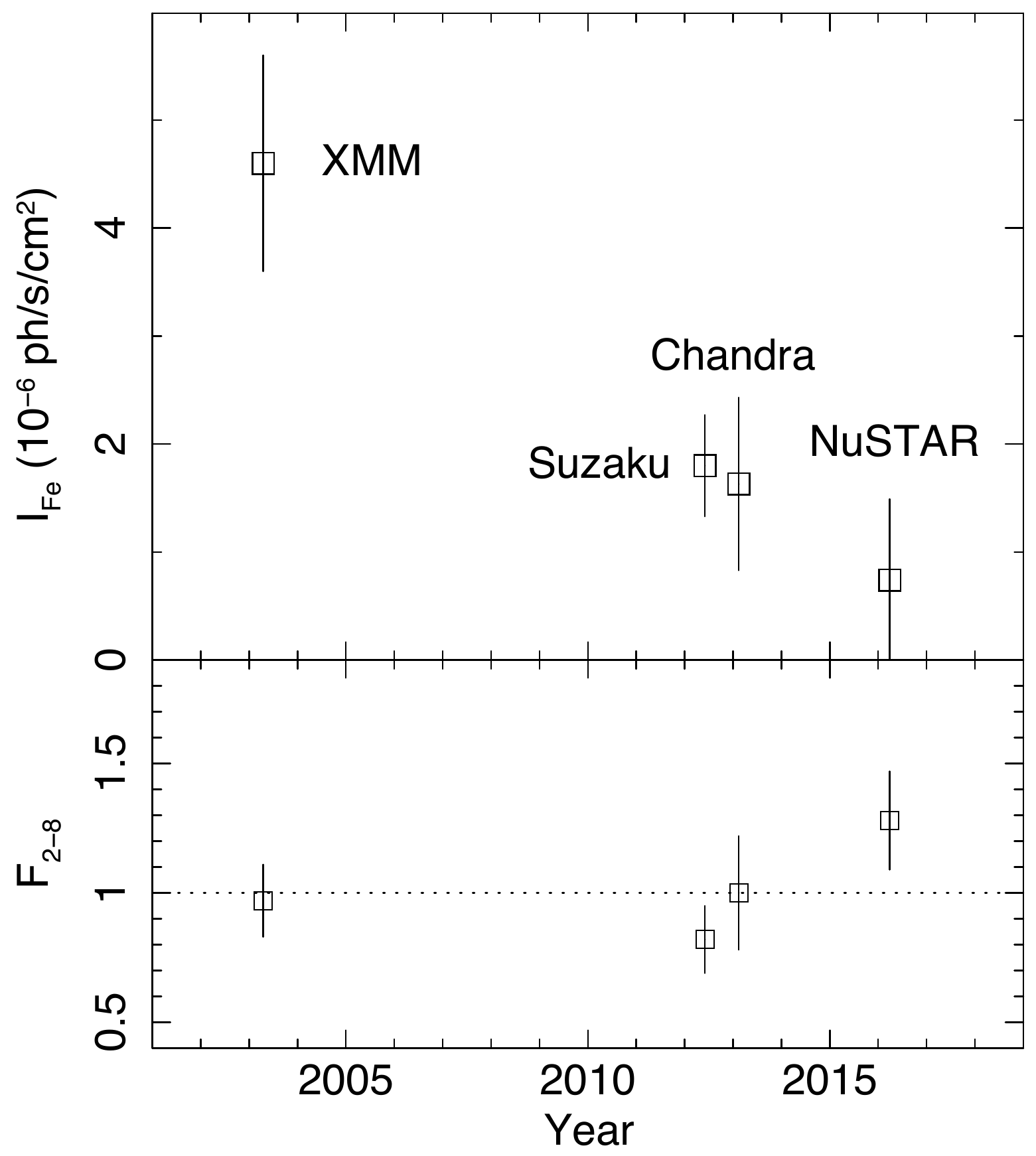}}
\caption{Light curves of Fe K line (a: Upper panel) and the 2-8 keV
  continuum (b: Bottom panel) of IRAS F00183--7111, observed with four
  X-ray observatories over 2003-2016. The continuum flux is in units
  of $10^{-14}$ \ergpspsqcm. Note that the joint fit determines the
  continuum shape better than in an individual fit, helping reducing
  the uncertainties of the line fluxes. The line flux appears to
  decrease continuously by factor of $\sim 4$ since the first
  observation with XMM-Newton while the continuum flux remains at
  comparable level within 30\%.}
\end{figure}

The Fe K line is however not detected in the NuSTAR data. We believe
that the lower resolution of the detector (0.4 keV in FWHM) is
unlikely to be the reason for the non detection, since it is
compatible with the line width measured above and would instead be
optimal for a detection. The Fe K band data with, say, 200-eV
intervals, which moderately oversample the detector resolution reveal
no excess emission at around 5 keV while it would be easily
detectable if the line flux remains the same. The 90\% upper limit of
line intensity is $1.9\times 10^{-6}$ \phpspsqcm, if the line energy
and width of the XMM, Chandra and Suzaku combined spectrum are
assumed. A possible line-like excess (unresolved) is instead seen at
5.4 keV ($7.28\pm 0.15$ keV in rest energy) with an intensity of
$(1.2\pm 0.5)\times 10^{-6}$ \phpspsqcm, but its detection is
uncertain ($\sim 2\sigma$). The apparent lack of the Fe K feature
prompted us an investigation of line variability, as the X-ray
observations of this ULIRG span over 13 yr (Table 1).

We performed a joint fitting of a Gaussian to the Fe K feature again,
but the line intensity was left independent between different
observations. The continuum was modelled by an absorbed
power-law and its normalization was left independent likewise (see
Sect. 3.2.2) while power-law slope and absorbing column of
cold absorption were kept common between the observations. We used the
3-20 keV data from NuSTAR and 2-8 keV data from the other instruments
jointly (Fig. 3). The photon index and absorbing column density as measured in
the galaxy rest-frame are found to be $\Gamma = 2.3\pm 0.2$ and \nH
$=(1.3\pm 0.3)\times 10^{23}$ \psqcm. The line centroid and the width,
common to all the observations, are found identical to those reported
above. The line intensity variability is plotted in Fig. 4a. The line
flux appears to decline monotonically since the first observation with
XMM-Newton in 2003. The line flux observed with Suzaku and Chandra in
2012-2013 is about half the flux measured by XMM-Newton and the NuSTAR
measurement is further down. With the four data points, the hypothesis
of linearly declining line flux, for example, is favoured by {\sl F} test to a
constant line flux at $\sim 98$\% confidence. However, we have only
four measurements, three of which are close together. The putative
line variability largely relies on the high flux measured by
XMM-Newton. While the consistently strong line fluxes measured with
the two different detectors of XMM-Newton, the pn and MOS cameras,
support the reliability of the strong XMM-Newton flux, another data point
after the NuSTAR observation with good quality is desirable to verify
the trend of fading Fe line.

\subsubsection{2-8 keV continuum}

In contrast to the Fe K emission, the neighbouring continuum remains
at similar flux level (Fig. 4b).  Fluxes of the four observations
derived from the joint fit in which the power-law normalizations were
left independent between the observations are used to plot Fig. 4b.

All the four measurements agree within 30\%. Given the statistical
uncertainties of individual measurements of $\sim 15$-20\% and the
general cross-calibration error of $\sim 10$\% (Ishida et al. 2011,
Kettula, Nevalainen \& Miller 2013; Madsen et al 2015), no significant
variability in continuum flux can be observed. The most deviated
NuSTAR flux is $\sim 30\pm 20$ \% above the mean value of the other
three. Even if this real, its variability is contrary to the declining
Fe K line flux. 

\subsubsection{NuSTAR Hard X-ray spectrum}

\begin{figure}
\centerline{\includegraphics[width=0.4\textwidth,angle=0]{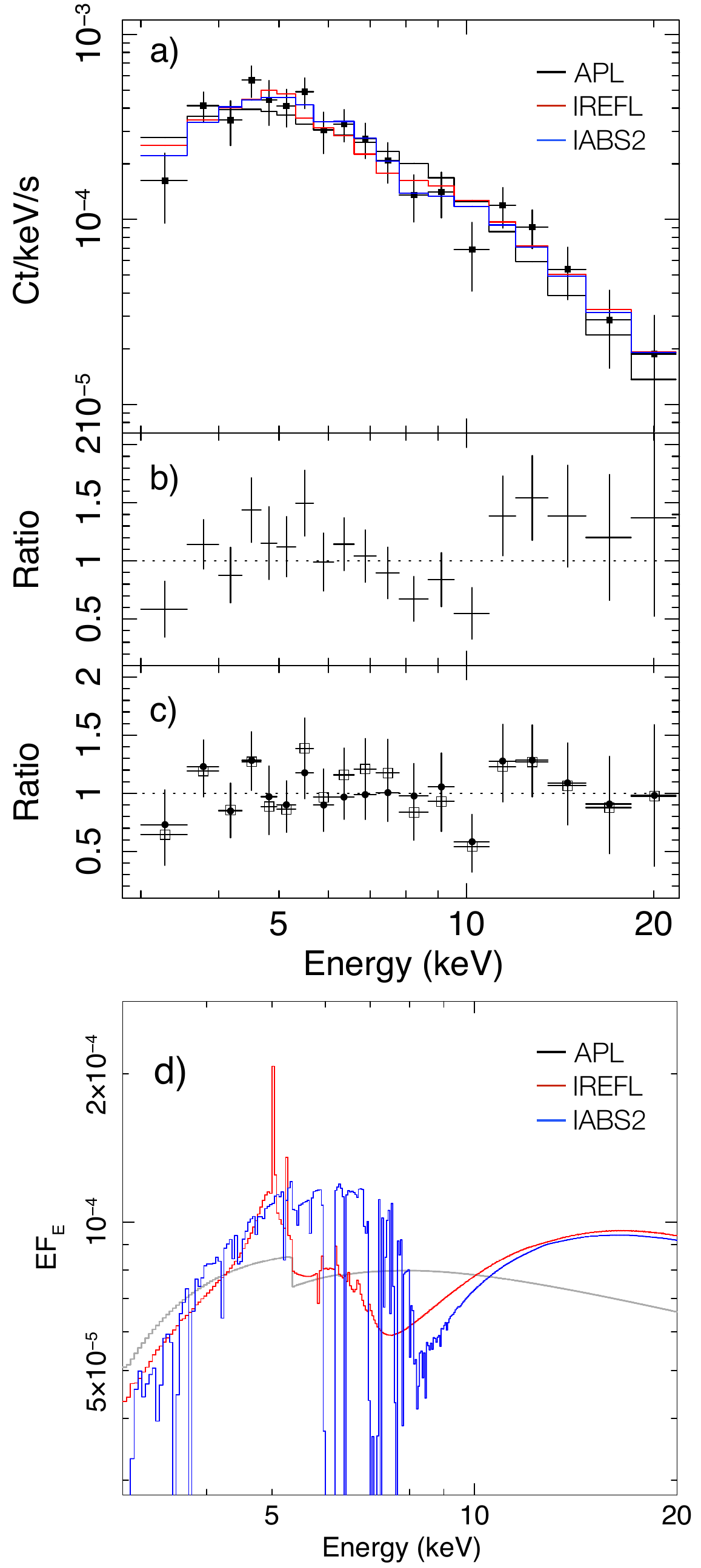}}
\caption{a) The NuSTAR count rate spectrum of IRAS F00183--7111 in the
  3-20 keV band. The three solid-line histograms plotted along the
  spectral data show best-fit models of the absorbed power-law (black,
  APL in Table 2) and the reflection spectrum computed by {\tt
    xillver} (red, IREFL) and the ionized absorption spectrum (blue,
  IABS2). All the models are folded through the detector response; b):
  Data/model ratio for the APL model in a); c) for IREFL (open
  squares) and IABS2 models (filled circles); and d) The three models
  shown above, without response folding and plotted in flux units. }
\end{figure}

Here, we investigate the hard X-ray spectrum of IRAS F00183--7111
obtained from NuSTAR (Fig. 5) which has the sensitivity at energies
above 8 keV where the other instruments do not cover. With the
reflection scenario in mind (Sect. 1), the original aim of this NuSTAR
observation was to see whether a strongly absorbed component emerges
in the hard X-ray band. Apparently, there is no such an extra
hard-component above 8 keV (see Sect. 4.1 and Fig. 6 for details)
A question is then whether the observed hard X-ray emission is
reflected light of a totally hidden central source or its direct
radiation modified by moderate absorption, which we examine here using
the NuSTAR data.

The 3-20 keV data are binned so that each spectral bin has more than
14 net counts after background correction (Fig. 5a), and the $\chi^2$
minimisation was used to search for the best fits. The hard spectrum
measured with XMM-Newton, Chandra and Suzaku is presumably caused by
cold absorption of \nH $\sim 1\times 10^{23}$ \psqcm (Sect. 3.2.2). Since this
degree of absorption cannot be constrained well by the NuSTAR data
alone, because of the limited low-energy coverage (down to 4 keV in
the rest frame), we impose cold absorption with \nH $= 1.3\times
10^{23}$ \psqcm, as obtained from the joint fitting with the
low-energy instruments (Sect. 3.2.2), in the following spectral
analysis.

A simple power-law modified by the fixed cold absorption yields a
photon index of $\Gamma = 2.39\pm 0.16$ (Table 2). This fit leaves
systematic wiggles in the deviation of the data from the model as
shown in Fig. 5b, indicating possible presence of an Fe K absorption
edge, which corresponds to the broad deficit around 8 keV. A sharp
deficit at the observed energy of 10 keV may be an artefact due to
background (NuSTAR Observatory Guide 2014). Including an absorption
edge model, {\tt edge}, improves the fit and gives an edge threshold
energy of $\sim 7.5$ keV ($9.9\pm 0.3$ keV in the rest frame) with an
optical depth of $\tau = 0.9\pm 0.3$. The edge energy suggests highly
ionized Fe. However, a sharp, single edge, as described by the {\tt
  edge} model, is inappropriate for high-ionization Fe, as an
absorption spectrum would instead consist of a series of absorption
features from a range of ionization (Kallman et al 2004), which has to
be computed by photoionization models like XSTAR (Kallman \& Bautista
2001). Alternatively, such an absorption feature appears also strongly
in a reflection spectrum when reflection occurs in an optically thick,
ionized medium.  Below we test the two hypotheses of ionized
reflection and direct emission modified by an ionized absorber.

We first compare the data with the ionized reflection spectrum
computed with {\tt xillver} of Garc\'ia et al. (2013). A thick slab
assumed for reflecting medium, e.g. an accretion disc, in {\tt
  xillver} is an approximation to the relatively thick reflecting
medium. In fitting to the XMM-Newton, Chandra and Suzaku spectra, the
Fe line feature drives the fit and gives the ionization parameter of
the reflecting medium of log $\xi=3.2\pm 0.1$. Fitting it to the
NuSTAR data alone finds a higher value of ionization parameter, log $\xi =
3.5^{+0.3}_{-0.2}$, which has the reduced sharp line and the increased
Compton-broadened component, matching better the broad hump around 5
keV and the weak line (IREFL in Table 2, Fig. 5).

Secondly, the absorption model is tested, introducing an ionized
absorber to account for the high-ionization Fe K edge feature in
addition to the cold absorber. No Fe emission line is included. We
used the analytic XSTAR model {\tt warmabs} (Kallman 2016) to compute an ionized
absorption spectrum to compare with the data. Since no obvious
absorption lines of Fe {\sc xxv} and Fe {\sc xxvi} are visible in the
NuSTAR data, we assumed that these lines are too narrow to be resolved
and chose a small turbulent velocity of $v_{\rm turb} = 200$ km
s$^{-1}$.

Fitting the ionization parameter and the column density of the ionized
absorber gives log $\xi = 3.1\pm 0.2$ and $N_{{\rm H},i}=1\times
10^{24}$ \psqcm\ (Note the best-fit column density is just below the
maximum value computed for the model and the upper bound of the error
is not obtained.). When the ionized absorber is allowed to be
blueshifted (IABS2), as expected for a high-velocity outflow, the fit
improves with similar absorber parameters and the blueshift of
$-0.18^{+0.02}_{-0.01}\thinspace c$ (Table 2, Fig. 5).

The residuals of the fits with IREFL and IABS2 are shown in Fig. 5c to
compare with Fig. 5b from the absorbed power-law (APL). Those three
spectral models before folding through the detector response are shown
in Fig. 5d, illustrating that the Fe absorption-edge feature and its
shift to the higher energy is a key to match the data. Between the
reflection and absorption models, the (blueshifted) absorption model
gives a better fit than the reflection model but with more free
parameters. Given the fitting quality, we cannot say the absorption
model is strongly preferred over the reflection model, e.g. a test by
Bayesian information criterion (BIC, Schwarz 1978) favours the
absorption model but the difference in BIC ($\Delta {\rm BIC}=2.6$)
indicates its preference is positive but not sufficiently strong (Kass
\& Raftery 1995). A critical test would be a detection of the Fe
absorption lines with higher-resolution data. The CCD resolution,
e.g. FWHM$\sim 150$ eV, would suffice if they exist. We returned to
the XMM-Newton, Chandra and Suzaku data to see whether one of the
absorption features due to Fe {\sc xxv} at the observed energy of 6
keV is present. There is marginal ($2\sigma $) evidence of an
absorption line at $6.07\pm 0.05$ keV (the rest-frame $8.05\pm 0.07$
keV, see Fig. 2), when all the datasets are jointly fitted by a
Gaussian. The line is unresolved (the 90\% upper limit of the
dispersion is 0.35 keV) and the intensity is $(-3.2\pm 1.6)\times
10^{-7}$ \phpspsqcm, corresponding to the equivalent width with
respect to the neighbouring continuum is EW $=-0.16\pm 0.08$ keV. The
IABS2 model obtained for the NuSTAR data, adjusted to the continuum
level and with the Gaussian for the Fe K emission, is shown in
Fig. 2. The data are compatible in the line energy and its depth
predicted by the absorption model. The unresolved line is consistent
with the small turbulent velocity chosen for the model. Albeit the
detection is inconclusive, the possible absorption line is found where
the blueshifted absorption model predicts.

\begin{table}
\begin{center}
\caption{NuSTAR spectrum of IRAS F00183--7111}
\begin{tabular}{lcccr}
Model & $\Gamma $ & log $\xi$ & log $N_{\rm H,i}$ & $\chi^2$/dof \\
(1) & (2) & (3) & (4) & (5) \\[5pt]
APL & $2.32^{+0.16}_{-0.16}$ & --- & --- & 22.5/17 \\ 
IREFL & $2.08^{+0.11}_{-0.12}$ & $3.5^{+0.3}_{-0.2}$ & --- & 15.0/16 \\
IABS1 & $2.38^{+0.17}_{-0.18}$ & $3.1^{+0.2}_{-0.3}$ & $24^{+0}_{-0.4}$ & 19.8/15 \\
IABS2$^{\dag }$ & $2.45^{+0.15}_{-0.15}$ & $3.0^{+0.1}_{-0.2}$ & $24^{+0}_{-0.2}$ & 9.6/14 \\
\end{tabular}
\begin{list}{}{}
\item[] Note --- The NuSTAR 3-20 keV spectrum was fitted with various
  models. (1) Fitted models, APL: a power-law modified by cold
  absorption with \nH $=1.3\times 10^{23}$ \psqcm, which is included
  in all the models; IREFL: an ionized reflection spectrum computed by
  {\tt xillver}; IABS1: a power-law modified by ionized absorption
  computed by XSTAR ({\tt warmabs} version 2.2); and IABS2: same as
  IABS1 except that a velocity shift of the absorber is
  permitted($^{\dag}$the best fitted value indicates a blueshift of
  $-0.18^{+0.02}_{-0.01}\thinspace c$). (2) Photon index; (3)
  logarithmic value of the ionization parameter $\xi $ of reflecting
  medium or absorber in units of erg cm s$^{-1}$; (4) logarithmic
  value of column density of the ionized absorber $N_{\rm H,i}$ in
  units of \psqcm; (5) $\chi^2$ over degrees of freedom.
\end{list}
\end{center}
\end{table}

\section{Discussion}

\subsection{Nuclear obscuration}

\begin{figure}
\centerline{\includegraphics[width=0.37\textwidth,angle=0]{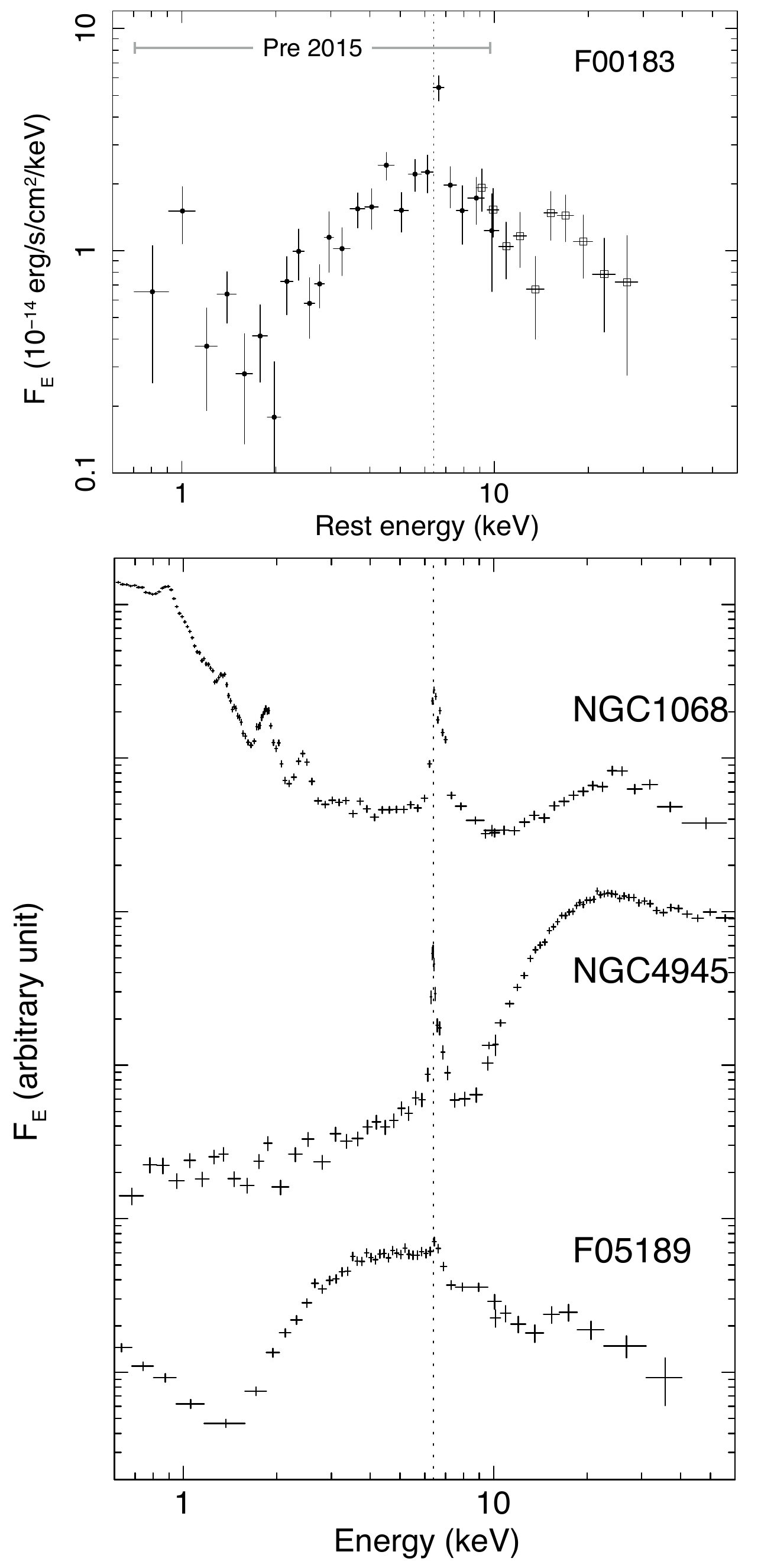}}
\caption{The broad-band X-ray spectra of IRAS F00183--7111 and three
  obscured AGN with different degrees of absorption for a comparison:
  NGC1068, a (cold) reflection-dominated spectrum without a
  transmitted component of the obscured nuclear source (\nH $\geq
  10^{25}$\psqcm); NGC4945, a reflection-dominated spectrum below 10
  keV with a transmitted component of the central source emerging
  above 10 keV (\nH $\sim 3.5\times 10^{24}$ \psqcm); IRAS
  F05189--2524, moderately absorbed nuclear source (\nH $\sim
  $(6-9)$\times 10^{22}$ \psqcm). The 0.6-8 keV data for IRAS
  F00183--7111 are obtained by combining the XMM-Newton, Chandra and
  Suzaku data. For the other three, XMM-Newton data are used for the
  0.6-10 keV band. Their spectra at higher energies are all obtained
  from NuSTAR. All the comparison spectra show Fe K emission peaking
  at 6.4 keV as indicated by the dotted line whereas that in IRAS
  F00183--7111 peaks at 6.7 keV (see Sect. 3.2.1). NGC 1068 and NGC 4945
  exhibit excess features in the NuSTAR band due to a cold reflection
  hump and an absorbed component, respectively.}
\end{figure}

The NuSTAR spectrum of IRAS F00183--7111 reported in this paper shows
no spectral hardening at high energies ($>8$ keV, as shown by the
spectral fits in Sect. 3.2.3), which would be observed if either
reflection from cold medium, expected from a Compton thick AGN, or
direct emission from a central source modified by a large absorbing
column of \nH $>10^{24}$\psqcm\ is present. This is illustrated by
Fig. 6, where the rest-frame broad-band X-ray spectrum of IRAS
F00183--7111 is compared with three obscured AGN with different
degrees of absorption. The spectrum of IRAS F00183--7111 is composed
of the 0.6-10 keV data, made from the XMM-Newton, Chandra and Suzaku
data, and the 8-20 keV NuSTAR data while those of NGC 1068 (Marinucci
et al 2016), NGC 4945 (Puccetti et al 2014) and IRAS~F05189--2524
(Teng et al 2015) are XMM-Newton and NuSTAR combined data. Observed
below 10 keV in NGC 1068 and NGC 4945 are reflected light only and
their spectra show upturns at energies above 10 keV due to a
reflection hump (Matt et al 1997) and a strongly absorbed continuum (\nH $\simeq
3.5\times10^{24}$ \psqcm), respectively. Such a spectral upturn lacks
in the hard-band spectrum of IRAS F00183--7111, which rather resembles
the moderately absorbed spectrum of IRAS F05189--2524. It leaves two
possible interpretations for the origin of the hard X-ray emission
from the ULIRG. Firstly, it could be reflected light of a hidden AGN
from a highly ionized medium, as originally suggested by Nandra \&
Iwasawa (2007) and Ruiz et al (2007). Direct emission from the central
source has to be totally suppressed by cold gas of an extremely large
absorbing column, e.g., \nH $\geq 10^{25}$\psqcm. Reflection off a
highly ionized medium would have a continuum spectrum similar to the
direct emission (e.g., Garc\'ia et al 2013), keeping the NuSTAR band
spectral slope steep, as observed. It naturally explains the strong Fe
{\sc xxv} line observed with XMM-Newton, Chandra and Suzaku. Secondly,
the hard X-ray emission could be of a moderately absorbed central
source which we see directly, as suggested by resemblance of the
overall spectral shape with that of IRAS F05189--2524 with the
absorption column density of \nH $\simeq (0.6-0.9)\times 10^{23}$
\psqcm (Severgnini et al 2001; Imanishi \& Terashima 2004, Ptak et al
2003, Grimes et al 2005, Iwasawa et al 2011, Teng et al
2015). Spectral fitting based on these two interpretations were made
against the NuSTAR spectrum in Sect. 3.2.3. Here we discuss their
likelihoods, combining with other information.

\subsubsection{Reflection scenario}

\begin{figure}
\centerline{\includegraphics[width=0.45\textwidth,angle=0]{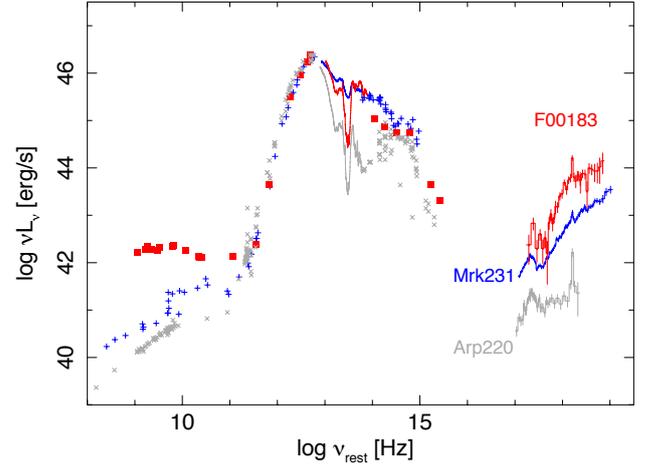}}
\caption{The radio to hard X-ray spectral energy distribution (SED) of
  IRAS F00183--7111 (in red). The radio data, submillimetre to
  infrared data, and the optical to UV data are taken from Norris et
  al (2012, references therein), Spoon et al (2009) and NED,
  respectively. For comparison, the SEDs of Arp 220 and Mrk 231,
  normalised to the far-infrared peak of IRAS F00183-7111, are
  overplotted in light grey and blue, respectively. This gives a
  convenient way to see the spectral differences between the three
  SEDs. The Spitzer IRS low-resolution spectra of the three sources
  are also included. Strong radio and X-ray excess emission from the
  active nucleus and the mid-IR emission from hot dust are evident in
  IRAS F00183--7111 and Mrk 231, relative to Arp 220. Note that Mrk
  231 is becoming more radio-loud in recent years than those the data
  plotted (Reynolds 2017).}
\end{figure}

Detection of a strong Fe line is generally considered to be good
evidence for a Compton thick AGN, as it leaves a reflection-dominated
spectrum below 10 keV. We refer to Nandra \& Iwasawa (2007) for
detailed discussion of the reflection scenario. Here, we look into a
few outstanding problems with this interpretation.

One peculiar feature in IRAS F00183--7111 (as a Compton thick AGN) is
the high-ionization (Fe {\sc xxv}) Fe line. Normally, the Fe K feature
in Compton thick AGN is dominated by a cold line at 6.4 keV, which
originates from a Compton-thick absorber itself. Even when any
high-ionization lines are present (e.g. in NGC 1068, Iwasawa et al
1997), they are minor components of the Fe K complex and the 6.4 keV
line is always the major feature. On the contrary, a cold Fe line is not
detected in IRAS F00183--7111 (see Sect. 3.2.1). In the reflection
scenario, the central source is assumed to be hidden from our direct
view by cold gas of an extreme thickness (thus invisible). Irradiation
of the obscuring gas by the central source however produces
reprocessed X-ray emission characterised by, e.g. a 6.4 keV Fe line,
and part of it enters our view, unless obscuration is very thick {\it and}
the coverage is complete. This may occur in nuclear sources of some
extreme objects like Arp 220 (Scoville et al 2017). However, in IRAS
F00183--7111, AGN-heated hot dust emission clearly visible in infrared
(Spoon et al 2004), for example, provides evidence against such an
extreme form of obscuration (see the SED comparison in Fig. 7 for the
strong contrast of mid-IR dust emission between IRAS F00183--7111 and
Arp 220). Thus the lack of reprocessed X-ray features from cold gas of
a putative Compton thick absorber is somewhat puzzling in the reflection
scenario.

Another critical issue is the line variability (Sect. 3.2.1). If the
line were indeed variable, the disconnected variability between the
line and continuum would give a problem with the reflection
scenario, since both originate in the same process and they should vary in
unison. A further line flux measurement with, e.g., XMM-Newton, will
provide a critical test against it.

The fit to the NuSTAR data (Sect. 3.2.3) shows that the reflection model
is still viable. In that case, the ionized reflecting medium has to be
sufficiently optically thick (\nH $\geq 10^{24}$ \psqcm) to produce
the edge absorption feature. The observed silicate depth might be
diluted by leaked IR continuum emission and the true column density of
cold clouds responsible for the silicate absorption could then be much
larger than the apparent value \nH $\simeq 1.7\times 10^{23}$
\psqcm\ (Spoon et al 2004) to provide a Compton-thick absorption
towards the central X-ray source.

\subsubsection{Direct emission scenario}

In the direct emission scenario, absorption by cold gas is moderate
(\nH $\simeq 1.3\times 10^{23}$ \psqcm ), but in addition to that, a
blue-shifted, highly ionized absorber with a large column \nH $\sim
10^{24}$ \psqcm\ is required to account for the absorption feature
around 8 keV observed in the NuSTAR spectrum (Sect. 3.2.3, Table 2). The
column density of the cold absorber is compatible with that inferred
from the silicate absorption (Spoon et al 2004). The high-ionization
absorber has an ionization parameter of log $\xi\sim 3$ and is only
detectable in X-ray. The inferred blueshift $v\simeq 0.18\thinspace c$
lies in the range of high-velocity outflows found in growing number of
Seyfert galaxies and quasars (e.g., Tombesi et al 2010). Given the
presence of powerful outflow signatures observed in mid-IR (Spoon et
al 2009), optical (Heckman et al 1990) and possibly radio (Ruffa et al
in prep.) bands, detection of X-ray outflowing gas comes as no
surprise. 

The 2-10 keV luminosity corrected for both cold and high-ionization
absorption is found to be $2.1\times 10^{44}$ \ergps. Discounting the
$\sim 14$\% starburst contribution (the star formation rate, SFR of
$220 M_{\odot}\thinspace {\rm yr}^{-1}$, Mao et al 2014; $230
M_{\odot}\thinspace {\rm yr}^{-1}$, Ruffa et al in prep.), we estimate
the AGN bolometric luminosity of IRAS F00183--7111 to be $3.3\times
10^{46}f^{-1}$ \ergps, where $f(<1)$ is a covering factor of obscuring
dust shrouds which absorb radiation from the central source and reemit
in infrared. This gives $L_{2-10}/L_{\rm bol,AGN}\sim 6\times
10^{-3}f$ (or the X-ray bolometric correction $k_{\rm bol}\sim 150
f^{-1}$). This value is smaller than that of typical AGN, e.g., $\sim
0.02$ for a normal quasar with a similar bolometric luminosity
(Marconi et al 2004), but comparable to those measured in local
U/LIRGs, including IRAS F05189--2524 (e.g. Imanishi \& Terashima 2004)
or in AGN accreting at a high Eddington ratio ($\lambda = L_{\rm
  bol,AGN}/L_{\rm Edd}$). We will return to the latter point later.

The absorption model fit applied to the NuSTAR spectrum ignores the Fe
emission-line observed with XMM-Newton, Chandra and Suzaku. Unless the
line-emitting gas is hidden from our view by some contrived change in
the nuclear obscuration, e.g. a passage of a Compton thick cloud
(Sanfrutos et al 2016), the Fe line strength needs to be explained by
the illumination of the central source. The observed mean line
luminosity is $(7\pm 2)\times 10^{42}$ \ergps. XSTAR predicts, with
the enhanced fluorescence yield of high-ionization Fe (e.g., Matt,
Fabian \& Ross 1996), compatible line luminosity can just be produced
by photoionization of thick medium of \nH $\sim 1\times 10^{24}$
\psqcm with log $\xi =3$ by the central source discussed above, under
favourable conditions, e.g. a large covering factor, without
suppression by the ionized absorber. Unlike the absorbing gas, the
line emitter has to be bound in the nucleus as it is observed at the
rest energy while their ionization states are similar. The line
emitter could be the accretion disc surface or the dense, base part of
the outflowing wind. Alternatively, it could also be optically thin
gas on a pc-scale. Although the line flux variability is not
conclusive (Sect. 3.2.1), its decline might be reverberation if a large
flare of the central source occured before the 2003 XMM-Newton
observation.


\subsection{Similarities to Mrk 231, high accretion rate}

Besides the high infrared luminosity, strong outflow signatures (Spoon
et al 2009; Heckman et al 1990) and the high-ionization gas inferred
from the X-ray Fe features hint that the black hole in IRAS
F00183--7111 may be accreting close to the Eddington limit
($\lambda\sim 1$), although there is no reliable way to measure its
black hole mass. If the above hypothesis of a moderately absorbed source
is correct, the nuclear source is relatively X-ray
quiet, which means a large X-ray bolometric correction, $k_{\rm bol}$,
or a steep optical/UV to X-ray spectral slope. In
the correlation diagram of $k_{\rm bol} - \lambda $ for the COSMOS
Type I AGN (Lusso et al 2010), $k_{\rm bol}$ of IRAS F00183--7111
corresponds to $\lambda\sim 1$.


We noticed that IRAS F00183--7111 shares interesting multiwavelength
properties with the well-studied local ULIRG/BAL quasar, Mrk 231. Mrk
231 exhibits strong galactic-scale outflows (Feruglio et al 2010;
Fischer et al 2010; Cicone et al 2012; Veilleux et al 2016) as well as
X-ray high-velocity winds (Feruglio et al 2015), has an X-ray quiet
nuclear source (log $(L_{2-10}/L_{\rm bol})\sim -3$), shows an Fe~{\sc
  xxv} emission-line (Reynolds et al 2017; Teng et al 2014), and has a
compact pc-scale radio source (Reynolds et al 2013). We note that the
$L_{\rm 2-10}/L_{\rm bol}$ estimated for IRAS F00183--7111 (Sect
4.1.2) becomes even closer to that of Mrk 231, if the covering factor
$f$ is smaller than unity. Mrk 231 was not as radio-loud as IRAS
F00183--7111 (Fig. 7) but is becoming more radio-loud in recent years
with elevated radio activity (Reynolds et al 2017). Curiously, the
mid-IR AGN tracer [Ne~{\sc v}]$\lambda 14.32\thinspace \mu $m is not
detected in either objects (Armus et al 2007; Spoon et al 2004,
2009). Contrary to the common wisdom for AGN being variable X-ray
sources, both IRAS F00183--7111 and Mrk 231 show stable intrinsic
brightness over years.

The black hole mass and the AGN bolometric luminosity of Mrk 231 have
been estimated in various methods but with large uncertainties. Among
the black hole mass measurements ranging from $1.3\times 10^7
M_{\odot}$ to $6\times 10^8 M_{\odot}$ (Tacconi et al 2002; Davies et
al 2004; Dasyra et al 2006; Kawakatu et al 2007; Leighly et al 2014)
and the AGN bolometric luminosity estimates of (0.4-1.1)$\times
10^{46}$ \ergps\ (Lonsdale et al 2003; Farrah et al 2003; Veilleux et
al 2009; Leighly et al 2014), we picked the respective medians
($M_{\rm BH}=8.7\times 10^7 M_{\odot}$ by Kawakatu et al 2007 and
$L_{\rm bol,AGN}=8.4\times 10^{45}$ \ergps\ by Leighly et al 2014) and
obtained an Eddington ratio of $\lambda = 0.76$ for Mrk 231. It
indicates that the black hole in Mrk 231 is likely operating close to
the critical accretion rate (e.g., Veilleux et al 2016). This may also
be supported by radio ejection events observed in the compact radio
source (Reynolds et al 2017), if analogy to the stellar mass black
holes (e.g., Fender, Belloni \& Gallo 2004) applies, as radio ejection
events occur only when they are accreting at $\lambda\sim 1$ in the
activity hysteresis. The production of the radio emission is thus not
of the ``radio mode'' in the low accretion-rate regime but of the
critical accretion, as seen in some radio-loud objects like 3C 120
(e.g. Ballantyne et al 2004).

The characteristic similarities mentioned above suggest that IRAS
F00183--7111 could also be a source of a high accretion rate. The
major divider between the two objects is the cold, line-of-sight
absorber which imprints the deep silicate absorption in the mid-IR
spectrum of IRAS F00183--7111. Instead, the weak silicate absorption
and the constraint on the jet angle ($<25^{\circ}$, Reynolds et al
2013) suggest that the inner nuclear structure in Mrk 231 is nearly
face-on. The AGN bolometric luminosity means a larger black hole mass
in IRAS F00183--7111: $M_{\rm BH}\sim 3\times 10^8 f^{-1} M_{\odot}$
for $\lambda = 1$. In fact, Type II AGN at $z\sim 2.5$ showing
high-ionization Fe K feature in the COSMOS field are found to be
hosted by Hyperluminous ($L_{\rm ir}\sim 10^{13}L_{\odot}$) IR
galaxies with their IR SED similar to IRAS F00183--7111 and their
nuclei are suspected to accrete close to $\lambda =1$ (Iwasawa et al
2012). Assuming both objects have a critical accretion disc, we
speculate the cause of the strangely stable X-ray luminosity observed
in the two sources might be a high optical depth of the thick disc
where the central source is located. Photon trapping starts to take
effects on an X-ray source around $\lambda\sim 0.2$ (e.g., Wyithe \&
Loeb 2012) and multiple Thomson scatterings would smear out their
intrinsic X-ray variability, although long term variability is
difficult to wipe out by this effect. Strong outflows observed in both
objects are a natural consequence of critical accretion discs
(e.g. Ohsuga et al 2002; Begelman 2012).

\begin{acknowledgements}
The scientific results reported in this article are based on
observations made by Chandra X-ray Observatory, Suzaku, NuSTAR and
XMM-Newton, and has made use of the NASA/IPAC Extragalactic Database
(NED) which is operated by the Jet Propulsion Laboratory, California
Institute of Technology under contract with NASA. Support for this
work was partially provided by NASA through Chandra Award Number
GO2-13122X issued by the Chandra X-ray Observatory Center, which is
operated by the Smithsonian Astrophysical Observatory for and on
behalf of the NASA under contract NAS8-03060. KI acknowledges support
by the Spanish MINECO under grant AYA2016-76012-C3-1-P and
MDM-2014-0369 of ICCUB (Unidad de Excelencia 'Mar\'ia de
Maeztu'). Support from the ASI/INAF grant I/037/12/0 – 011/13 is
acknowledged (AC, MB, EP, GL, RG and CV).

\end{acknowledgements}

\end{document}